\documentclass{article}

\usepackage{graphicx}
\usepackage{framed}
\usepackage{url}
\usepackage{authblk}

\begin{document}

\title{Image Watermaking With Biometric Data For Copyright Protection}

\author[1]{Barbier, Morgan\thanks{\texttt{morgan.barbier@ensicaen.fr}}}
\author[1]{Le Bars, Jean-Marie\thanks{\texttt{jean-marie.lebars@unicaen.fr}}}
\author[1]{Rosenberger, Christophe\thanks{\texttt{christophe.rosenberger@ensicaen.fr}}}

\affil[1]{ENSICAEN-UNICAEN-CNRS, GREYC, F-14032 Caen, France}

\maketitle

\begin{abstract}
  In this paper, we deal with the proof of ownership or legitimate usage of a digital content,
  such as an image, in order to tackle the illegitimate copy. The proposed
  scheme based on the combination of the watermarking and cancelable
  biometrics does not require a trusted third party, all the exchanges
  are between the provider and the customer. The use of cancelable biometrics permits to provide a privacy compliant proof of identity.  We illustrate the
  robustness of this method against intentional and unintentional
  attacks of the watermarked content.
\end{abstract}

\vspace{0.5cm}

\vspace{0.5cm}

\section{Introduction}
\label{sec:in}
In the last decade, intensive work led to different methods to manage
the author rights of images using cryptographic protocols
\cite{DRMThesis,DRMMohanty}. However, these schemes ensure a low
security guarantee of the data owner. Indeed, these systems cannot link
the owner identity to its use rights, this is intrinsic to
cryptography where the security is related to the knowledge of a
secret, not to an identity. In order to tackle this problem, some
researchers thought to use biometric systems. The embedding of
biometric data into an image has been proposed for the first time in
2004 to link the photographer iris to taken pictures
\cite{SDCFridrichBlythe}. The same technique was also proposed for the
purpose of multi-biometric applications consisting in embedding
fingerprint into visage pictures \cite{MultiBio, MultiBio2}, but not
any to prove the ownership of an image. The previous solutions embed
very sensitive data into a shared image, which affect hugely to the
user privacy and appears a significant problem. Recent schemes have been 
proposed in order to obtain cancelable biometric data
\cite{Teoh04}. This process is called BioHashing.  The knowledge of
this data does not provide any information of the original biometric data,
which ensures the user privacy. However, it implies to be carefull
with the seed initialization \cite{lacharme2013reconstruction}. Indeed, if an attacker
knows both BioCode and this seed, he can forge a pretty a good estimate of the FingerCode.\\

In this article, we present a new scheme of proof of ownership
exploiting cancellable biometrics. With the help of cryptographic
protocols, the proposed method ensures the watermark security and preserve the
owner privacy. A particular value is shared between these two actors
and can be seen as a biometric commitment. As far we know, it is the
first time that a biometric commitment is done, and this pretty
property avoids a trusted third party; which is a significant improvement for
some applications. In this setting, this article is a direct
improvement of \cite{barbier2014tatouage}.\\

First, we define the requirements that a such
system has to respect for the different actors in terms of security and privacy. Second, we detail the proposed method. Experimental results show the benefit of this solution. We perform in the following
a security and privacy analysis, where we discuss about the respect of
the previous requirements. Finally, we conclude this paper and give different perspectives of this study.

\section{Security and privacy requirements}
\label{sec:Secu}

A biometric system handles, by definition, very sensitive personal
data of users. These data aim to be protected for prevent their
usurpation, modification or falsification. In our context, we
define two main actors. The first one is the \textbf{customer} who
wishes to acquire the copyright of a digital data proposed by a
\textbf{provider} or \textbf{owner}. Afterwards, the provider should be able to check that a
customer has the copyright for a particular data. Moreover, a customer should be able to prove he personally had previously obtained the right to use this digital content. In this context, we
introduce the main security and privacy requirements of the previous
system:
\begin{itemize}
\item[$R_1$:] \textbf{Proof of ownership} assures that the legitimate
  customer can prove, at any moment, his right to use the data.
\item[$R_2$:] \textbf{Proof of paternity} assures that the provider
  can prove, at any moment, its paternity of the digital content.
\item[$R_3$:] \textbf{Unlinkability} of the watermarks for a same
  user in different data. An attacker should not be able to link the
  different watermarks of a customer.
\item[$R_4$:] \textbf{Confidentiality of customer data} ensures to
  not be known by anybody.
\item[$R_5$:] \textbf{Confidentiality of provider data} ensure to
  not be known by anybody.
\item[$R_6$:] \textbf{Customer data sovereignty} implies that the
  copyright checking can be done only with the agreement of the
  customer.
\item[$R_7$:] \textbf{Provider data sovereignty} implies that the
  copyright checking can be done only with the agreement of the
  provider.
\item[$R_8$:] \textbf{Non-falsification} prevents a
  customer/attacker to make a legitimate mark.
\item[$R_9$:] \textbf{Non-repudiation} prevents a provider to deny
  the sale of a media.
\end{itemize}

\begin{figure*}[!ht]
  \centering
  \includegraphics[width=11cm]{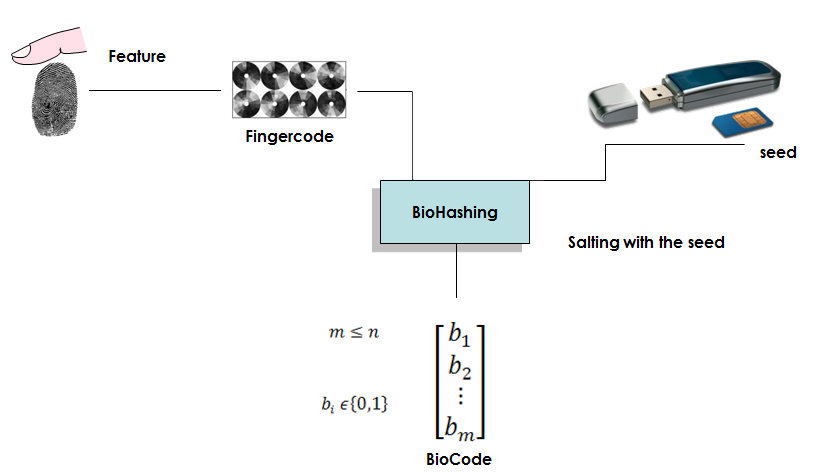}    
  \caption{Use of the BioHashing algorithm}
  \label{pictures:biohashing}
\end{figure*}

\section{Proposed method}
\label{sec:méthode}

The proposed method has for objective to protect at the same time the ownership of an image and the right of use for a customer. As biometrics is the only technology that really guarantees the user identity, we use this kind of information in the proposed method. We embed a mark into an image in order to prove these two aspects. The watermark is computed from a cancelable biometric data to permit a secure and privacy compliant identity verification. We detail in the next section the used watermarking method.

\subsection{Image watermarking}

\begin{figure}[h!]
$$
\begin{array}{ccc}
\includegraphics[width=2.7cm]{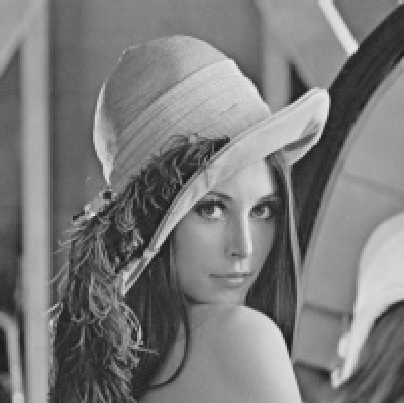}
~\includegraphics[width=2.7cm]{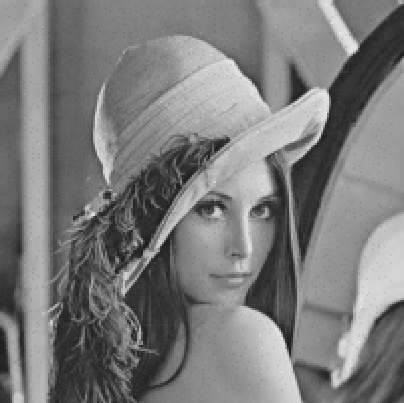}
~\includegraphics[width=2.7cm]{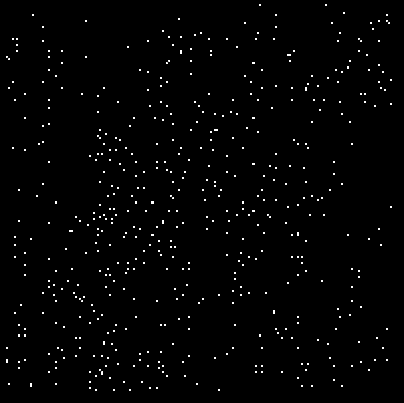}
\end{array}
$$
\caption{From left to right: original image, watermarked, difference (multiplied by 100)}
\label{pictures:illustration}
\end{figure}

In the proposed approach, we use the watermarking method introduced by Wenyin
and Shih \cite{Wenyin2011}. This method uses image texture parameters
so-called Local Binary Patter (LBP), to select pixels to embedded the
mark. It ensures a good robustness to different distortions like
compression and cropping simulating intentional
and unintentional attacks of the watermarked image. Figure \ref{pictures:illustration} an example of watermarked image with this approach, the difference between the original and the watermarked images multiplied by 10 is displayed as illustration.

\subsection{Image identifier computation}

\begin{figure}[!h]
    \centering
    \includegraphics[width=7cm]{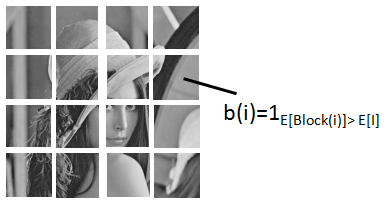}    
    \caption{Image identifier computation}
    \label{pictures:identifier}
\end{figure}

The copyright protection must be verified easily and be related to the image. We propose to define an image identifier that is easy for anybody to compute. The image is divided in blocks where the number is related to the size in bits of the image identifier. As for example, Figure \ref{pictures:identifier} shows the method to compute an identifier with 16 bits. We compute first the average gray level value called $E[I]$ where $I$ is the image. Second, we compute the average grey level value of each block (denoted as $E[Block]$) and we compare it to $E[I]$. Based on the result, we assign a value 1 or 0 to each of the bit of the image identifier.

The BioHashing algorithm is used to compute the cancellable biometric data, it is detailed in the next section.

\begin{figure*}[!htbp]
  \centering
  \includegraphics[width=12cm]{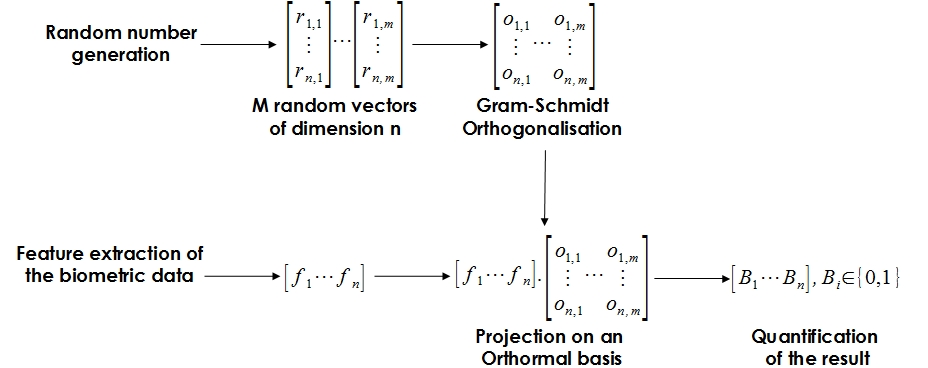}    
  \caption{Principle of the BioHashing algorithm}
  \label{pictures:biohashing2}
\end{figure*}

\subsection{BioHashing algorithm}
\label{sec:hash}

The BioHashing algorithm transforms a real-valued vector of length $n$
(i.e. the FingerCode, resulting from a feature extraction method) into
a binary vector of length $m \leq n$ (i.e. the BioCode), as first
defined by Teoh \emph{et al.} in \cite{Teoh04}.\\

It consists in projecting the FingerCode on an orthogonal basis
defined by a random seed (considered here as a secret), to generate
the BioCode. The template transformation uses the following algorithm,
where the inputs are the random seed and the FingerCode $F$; the
output is the BioCode $B$:
\begin{enumerate}
\item For  $i=1,\ldots ,m$, $m  \leq n$ pseudorandom vectors  $v_i$ of
  length $n$ are generated (from  the random seed) and are gathered in
  a pseudorandom matrix.
\item The Gram-Schmidt  algorithm is applied on the  $m$ vectors $v_i$
  of  the  matrix,  for  the  generation of  $n$  orthonormal  vectors
  $V_1,\ldots ,V_m$.
\item For  $i=1,\ldots ,m$,  $m$ scalar products  $p_i =  <F,V_i>$ are
  computed using  the FingerCode $F$  and the $m$  orthonormal vectors
  $V_i$ .
\item The  $m$-bit biocode $B=(B_0,\ldots ,B_m)$  is finally obtained thanks to 
the following quantization process:
$$B_i= \left\{
  \begin{array}{l}
    0 \quad \mbox{if} \quad p_i<t  \\
    1 \quad \mbox{if} \quad p_i\geq t,\\
  \end{array}
\right.$$
where $t$ is a given threshold, generally equal to $0$.
\end{enumerate}

When used for authentication the {\it Reference BioCode} (computed
from the FingerCode after enrollment and after exhibiting the secret)
is compared with the {\it Capture BioCode} (computed from the
FingerCode computed after a new capture with the secret) with the
Hamming distance. If this value is lower than a specified threshold
set by the system administrator, the identity of the user is
checked. Roughly speaking, the first part of the algorithm, including
the scalar products with the orthonormal vectors, is used for the
performance requirements and the last step of the algorithm is used
for the non-invertibility requirements of the BioHashing algorithm. As
mentioned before, the random seed guarantees the diversity and
revocability properties.\\

The user authentication protocol applies multiple times the BioHashing
algorithm.

\subsection{Mark insertion}

\begin{figure*}[ht!]
    \centering
    \includegraphics[width=12cm]{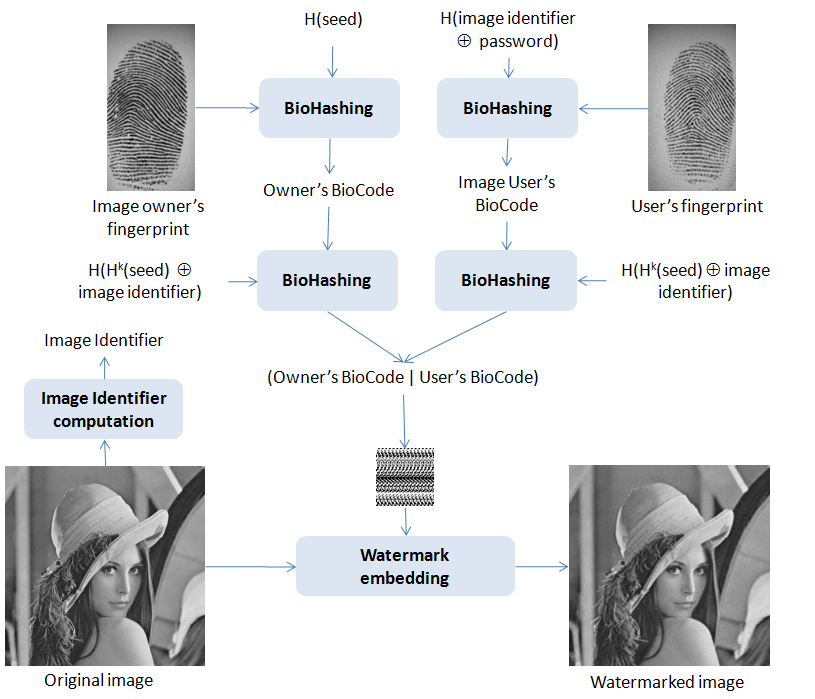}    
    \caption{Inserting the mark in an image}
    \label{pictures:insertion}
\end{figure*}

Figure \ref{pictures:insertion} describes the proposed method to
insert in an image some information related to the image's owner and
customer. The mark we embed is defined as the repetition the global
Biocode (16 times of a 256 bits BioCode to generate the 64x64 pixels
to fulfill the requirements of the used watermarking method). The seed
is a secret defined by the owner. The value $H(H^k(seed) \oplus
~image~identifier)$ can be seen as a cryptographic commitment shared
between the image's owner and customer. Each time a BioCode is issued, the
value of $k$ is incremented (to guarantee different commitments for
different applications of the watermarking approach). The term $H^k(seed)$ corresponds to an OTP (dynamic value) generated from a seed. The use of the
BioHashing algorithm ensures to not be able to obtain the data at a
lower level as it is a non invertible function. Note that the customer should provide his/her fingerprint (as identity proof) and a password (that can be the same for different images).\\

The embedded mark in the image biometric data from both the image's
owner and user (concatenation of two BioCodes). The verification of
property or right to use is possible by considering fingerprint as
identity proof for owner or user side. Note that at the first level,
the owner's BioCode is not related to the image as the right of user
could be provided multiple times. For user side, the BioCode is
related to the image as an user will buy only once the right of use for a
specific image.

\subsection{Mark verification}

If the owner finds its image on internet, the mark can be extracted
(if present). First, based on the image identifier that can be computed
from this image, the owner can look at existing BioCodes for this
image. If this BioCode exists in his/her database, the $k$ value is
determined. Second, the right of use must be checked. To achieve this
goal, the webmaster of the website where the image is used can be
contacted in order to verify this point. In this case, the user's
fingerprint should be given and the user's BioCode generated (with the
$k$ value set previously). If the Hamming distance between this
BioCode and the one present in the image matches is lower a certain
threshold (set by the owner), the user's right of use is validated.\\

The proposed method permits to verify at the same time the ownership
and the right of use. An attacker is not able to generate a valid
user's BioCode as he/her does not know the cryptographic commitment.

\section{Experiments}
\label{subsec:perf}

In this section, we analyze first the robustness the watermarking method face to attacks. Second, we analyze the performance of the biometric recognition (owner or user side) even if the watermarked image has been altered. In the next section, we define the experimental protocol used in this study.

\subsection{Protocol}

\begin{figure}[h!]
$$
\begin{array}{ccc}
\includegraphics[height=4 cm]{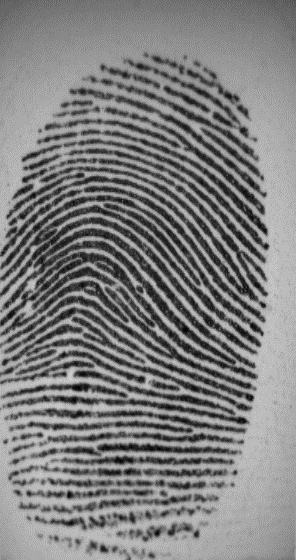} & \includegraphics[height=4 cm]{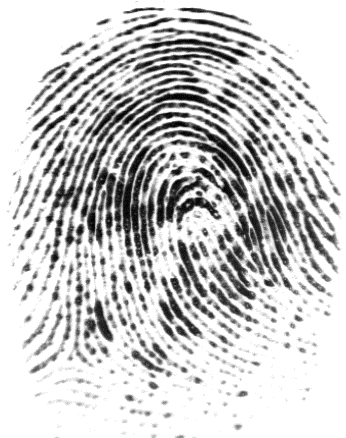} &
\includegraphics[height=4 cm]{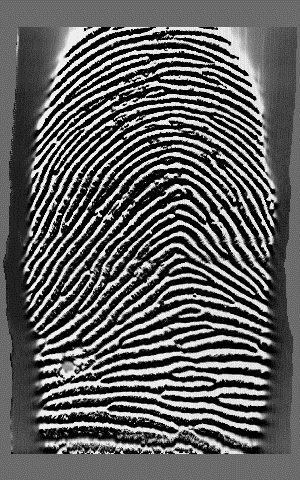}\\
(a) & (b) & (c) 
\end{array}
$$
\caption{One fingerprint example from each database: (a) FVC2002 DB2, (b) FVC2004 DB1, (c) FVC2004 DB3}
\label{database}
\end{figure}

In this study, we used three fingerprint databases, each one is composed of 800 images from 100 individuals with 8 samples from each user:
\begin{itemize}
\item FVC2002 benchmark database DB2: the image resolution is $296 \times 560$ pixels with an optical sensor "FX2000" by Biometrika ;
\item FVC2004 benchmark database DB1: the image resolution is $640 \times 480$ pixels with an optical Sensor "V300" by CrossMatch ;
\item FVC2004 benchmark database DB3: the image resolution is $300 \times 480$ pixels with a thermal sweeping Sensor "FingerChip FCD4B14CB" by Atmel.
\end{itemize}
~\\
Figure \ref{database} presents one image from each database. We can see that fingerprints are quite different and representative of the different types of fingerprint (acquired with sensors using different technologies).\\

As FingerCode, we used Gabor features (GABOR) \cite{Manjunath1996} of size n=512 (16 scales and 16 orientations) as template. These feature are very well known and permit a good texture analysis of a fingerprint. For each user, we used the first FingerCode sample as reference template. Others are used for testing the proposed scheme. BioCodes are of size m=256 bits. In order to quantify the performance of the proposed approach, we computed 1400 comparisons (with the Hamming distance) between the BioCode embedded in the watermarked image and the computed one for each user.\\

\begin{figure*}
  \centering
  \includegraphics[width=12cm]{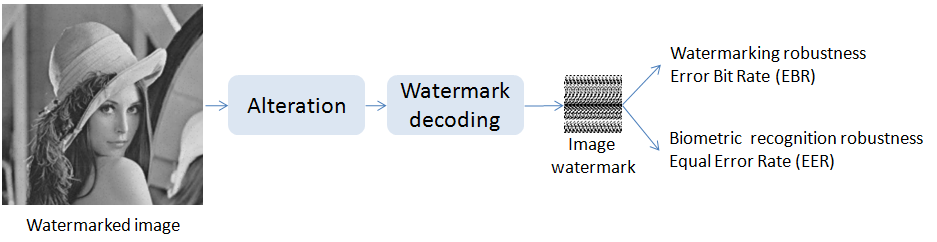}    
  \caption{Evaluation process of the proposed approach}
  \label{pictures:evaluation}
\end{figure*}

\begin{figure*}[!ht]
  \centering
  \includegraphics[width=12cm]{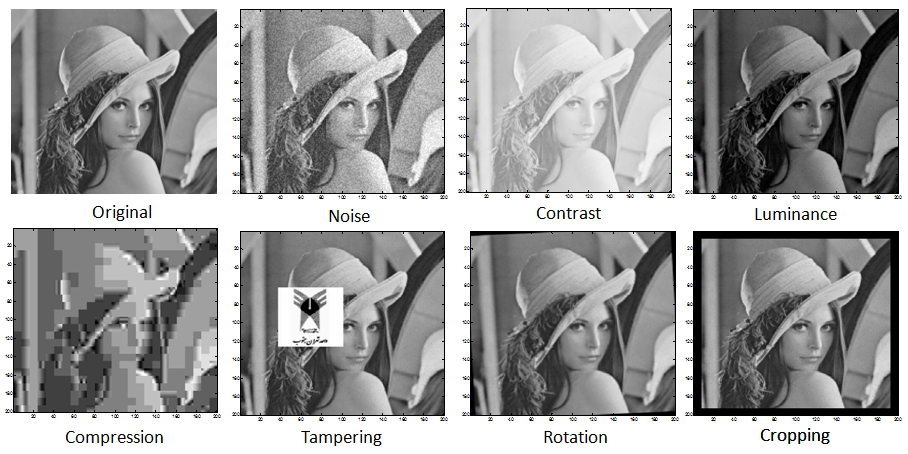}    
  \caption{Alterations of the watermarked image}
  \label{pictures:alterations}
\end{figure*}

The evaluation process is detailed in Figure \ref{pictures:evaluation}. We apply many alterations on the watermarked image (illustrated on Figure \ref{pictures:alterations}) simulating active or passive attacks.\\

The robustness of the watermarking algorithm is estimated for the Error Bit Rate metric (EBR)
defined as follows. The performance of the biometric recognition is determined by the Equal Error Rate (EER). This metric computes the performance of the biometric system when the threshold is set to have a compromise between the false acceptance rate and false rejection rate.

$$EBR=\frac{\sum \sum C(x,y)\oplus\tilde{C}(x,y)}{M.N}$$  

Where $C(x,y)$ is the initial value of the mark at pixel $(x,y)$, $\tilde{C}$ is the decoded mark, M and N are respectively the number of lines and columns of the mark (in our case, N=M=64), the symbol $\oplus$ corresponds to a logical XOR.

\section{Results}

\begin{figure}[!ht]
  $$
  \hspace{-1cm}
  \begin{array}{cccc}

    \includegraphics[width=4cm]{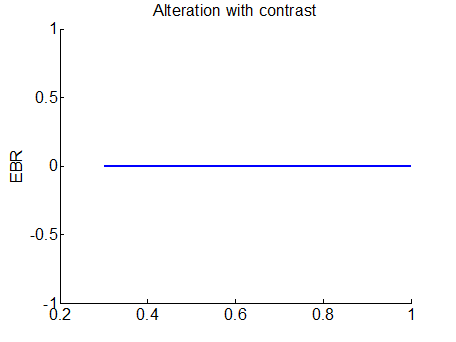}
    \includegraphics[width=4cm]{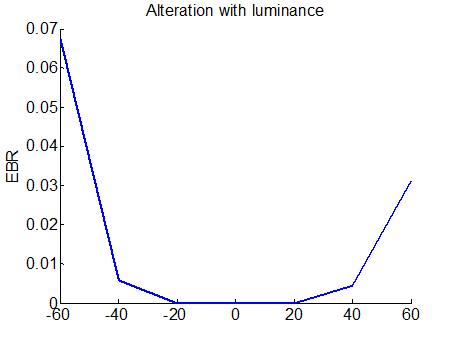}
    \includegraphics[width=4cm]{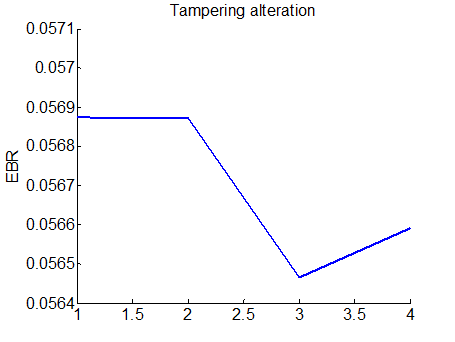}\\
    \includegraphics[width=4cm]{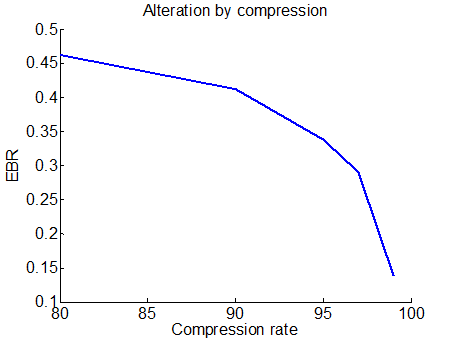}
    \includegraphics[width=4cm]{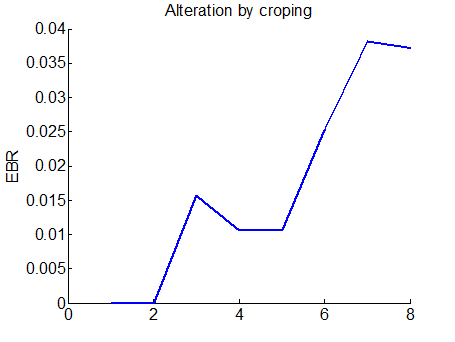}
    \includegraphics[width=4cm]{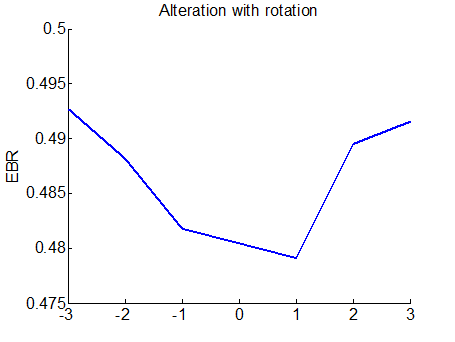}\\
    \includegraphics[width=4cm]{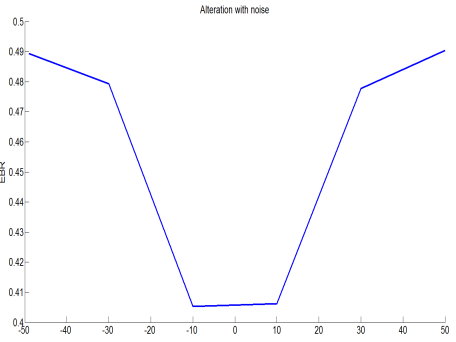}
  \end{array}
  $$
  \caption{EBR evolution for different distortions}
  \label{pictures:performance_alterations}
\end{figure}

First, we study the robustness othe image watermarking method face to different alterations illustrated in Figure \ref{pictures:alterations}. Figure
\ref{pictures:performance_alterations} presents the evolution of the EBR for each alteration. We can see that the watermarking method is completely invariant to the contrast alteration. Some alterations have as impact a low modification of the mark such as the luminance, cropping or tampering. Other alterations have  strong impact of the mark.\\

\begin{figure*}
    \centering    
    \includegraphics[width=12cm]{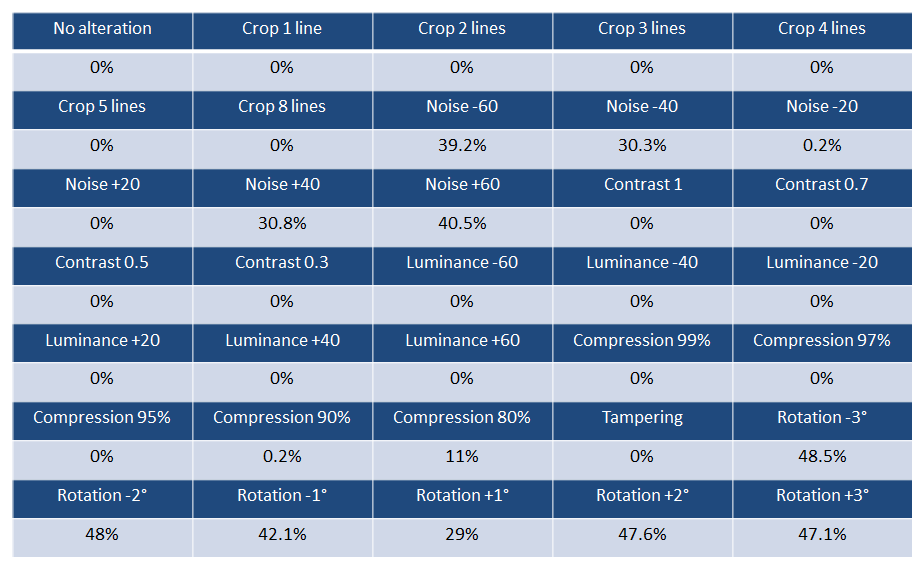}    
     \caption{Biometric recognition performance for the FVC2002 DB2 dataset. Results are similar for the two other datasets.}
    \label{pictures:performance}
\end{figure*}

Figure \ref{pictures:performance} gives the biometric recognition performance (of the owner or user) when the watermarked image is attacked considering different alterations. We can see that for invisible alteration, the EER value is near 0\%. For more visible alterations, EER values can be very high but there is no great benefit for an attacker to be able to suppress the copyright with such an alteration of the image. These results shows the robustness of the proposed approach (that can be of course improved).

\section{Security and privacy analysis}

\subsection{Attack model}
We consider the strongest attack model where the 
attacker is able to modify data. However, its actions will be
limited thanks to using SSL channel to preserve communication
confidentiality, actor authentications and data integrity. Without
lost in generality, we may assume that the attacker is passive for
communications and active to attempt to create legal documents, to
usurp identities and to link the customer sales.

\subsection{Analysis}
In this section, we discuss about the proposed system and
the security and privacy requirements introduced in
Section~\ref{sec:Secu}.\\

\paragraph{$R_1$}
In order  to prove that a  customer is legitimate, the  court begin to
check  the provider  BioCode, then  OTP value  is now  known,  and the
provider is  engaged on  this value. Finally,  the court  computes the
customer BioCode  with the hash of  the previous OTP.  If the provider
tries to cheat with the OTP  in order to harm the legitimate customer,
the provider BioCode computed cannot be checked.
 Moreover, our system is resilient to some modifications
carry out by the legitimate customer, as previously seen
Section~\ref{subsec:perf}. Thus as long as the customer
perform some modifications on its digital content listed in
Section~\ref{subsec:perf} like crop, contrast, luminance
and decent noise and compression; the customer will be recognized as
legitimate, so the requirement is respected.\\

\paragraph{$R_2$}
The first part of the embedded mark is dedicated only to the provider; this computation is based on the provider fingercode, some image characteristics and the seed. All these previous values can be supply by the provider, to proof the paternity of the digital content.\\

\paragraph{$R_3$}
Since the BioHashing is based on a projection of vector into a space
with less dimension, this operator is not bijective and is not
inversible. Moreover, for each digital content, the BioCode will be
completely different, then as far we know, it is impossible to make the
link from different marks to a customer; this requirement is
satisfied.\\

\paragraph{$R_4$}
To access to the customer data, an attacker has to inverse twice the BioHashing, which is impossible. However, estimations can be computed if and only if the seed is known; which is not the case up to customer's fingercode; $R_4$ is fulfilled.\\

\paragraph{$R_5$}
For the same argues that for the requirement $R_4$, we deduce that $R_5$ is satisfied.\\

\paragraph{$R_6$}
Since provider data used is only its FingerCode by the BioCode, 
it cannot be manipulated without its permission. The requirement
$R_4$ is fulfilled.\\

\paragraph{$R_7$}
For the same argues that for the requirement $R_6$, we deduce that $R_7$ is satisfied.\\

\paragraph{$R_8$}
The embedded mark is generated thanks to the provider's FingerCode and
an ID related to the digital content. Hence the marks
are totally different and an attacker cannot forge a legitimate
mark. The requirement $R_8$ is satisfied.\\

\paragraph{$R_9$}
Finally, if the provider suspect an user to use a digital content
without to be the right owner, the provider will take this user to
court. Then the justice will make the whole computation asking provider 
and customer's FingerCodes to tell the final decision.

\section{Conclusion and perspectives}

This paper contains a new method to achieve a proof of ownership of a
digital content with a biometric scheme preserving the privacy. The
resiliency of various distortions simulate an intentional and
unintentional attacks of the watermarked content shows a good
robustness of the introduced scheme for biometric checking.  Our main goal
was to avoid a trusted third party and then our scheme is entirely
satisfactory. \\

However, some parts of the scheme could improved. Indeed
further investigations will be conducted to make the watermark scheme
more efficient.  We may apply other watermark insertion algorithms and
replace the repetition code to a more adapted binary codes.  These
modifications should lead to a more robust scheme.  We could also
consider color image; since such a image contains more information,
we can expect to improve the robustness. 

\bibliographystyle{plain}

\small\bibliography{biblio}
\label{sec:biblio}
\end{document}